\documentclass[aps,prd,superscriptaddress]{revtex4}

\usepackage{amsfonts}
\usepackage{amsmath}
\usepackage{graphicx}
\usepackage{color}
\usepackage{amssymb}
\usepackage{esint}
\usepackage{empheq}
\usepackage{mathtools,leftidx}
\usepackage{float}
\usepackage[T1]{fontenc}
\usepackage{bbold}

\usepackage{soul}

\DeclareMathAlphabet\mathbfcal{OMS}{cmsy}{b}{n}

\begin{document}

\title{Spontaneous symmetry breaking in models with second-class constraints}

\author{C. A. Escobar}
\email{carlos.escobar@ciencias.unam.mx}
\affiliation{Departamento  de  F\'isica,  Universidad  Aut\'onoma  Metropolitana-Iztapalapa, San Rafael Atlixco 186, 09340 Ciudad de M\'{e}xico, M\'{e}xico}

\author{Rom\'an Linares}
\email{lirr@xanum.uam.mx}
\affiliation{Departamento  de  F\'isica,  Universidad  Aut\'onoma  Metropolitana-Iztapalapa, San Rafael Atlixco 186, 09340 Ciudad de M\'{e}xico, M\'{e}xico}

\begin{abstract}
In this work the spontaneous symmetry breaking in certain nonlinear theories with second-class constraints is explored. Using the Dirac's method we perform an analysis of the constraints and the counting of the degrees of freedom. The corresponding effective Hamiltonian is constructed explicitly. It is shown that on the surfaces where the effective Hamiltonian takes critical values the symplectic structure becomes degenerate. In particular, we demonstrate that under the condition of spontaneous symmetry breaking, which implies non trivial vacuum surfaces, second-class constraints behave as first-class ones on certain regions of the phase space, leading to undefined Dirac's brackets and to the modification of the number of degrees of freedom. As a physical consequence, these models can suffer from certain pathologies such as the existence of modes with an acausal propagation. Concrete examples where this phenomena occurs are described in detail.
\end{abstract}

\maketitle

\section{Introduction}

Symmetries play a fundamental role in the laws of nature. Modern physics, ranging from the Standard Model of particle physics to General Relativity, is grounded on symmetry principles and many physical phenomena can be well explained purely invoking invariance arguments. When a physical state, as the vacuum, is not invariant under the same symmetries as the theory that describe it, the initial symmetries of the Lagrangian are broken or slightly broken. In this case, notable physical phenomena such as phase transitions \cite{Lan1} or the Higgs mechanism \cite{Higgs1} arise via the spontaneous symmetry breaking (SSB) mechanism \cite{SSB2a}.

A model with SSB, via an explicit potential, usually includes nonlinearities otherwise a nontrivial vacuum cannot occur. This can lead to nonlinear constraints that may cause undesirable physical implications. A theory for a physical system must have a consistent mathematical and causal structure. It also should satisfy basic criteria such as having non-tachyonic propagating modes carrying positive energy. Moreover, it is said that the model possesses a well-posed Cauchy problem if its evolution is uniquely determined by smooth initial data satisfying the constraints, and/or if, under slightly initial data changes, the evolution varies continuously and respects the causal spacetime structure. On this matter,  the Cauchy-Kovalevskaya theorem ensures that, in globally hyperbolic spacetimes, any second order, linear, diagonal, hyperbolic (LDSH) system of differential equations has a well posed Cauchy problem \cite{Cauchy1}. However, when the corresponding equations are not an LDSH system of differential equations, there is no a general method to determine if  the theory has a well-posed Cauchy problem or not. Hence, one has to prove each physical property, as causality or positive definite energies, independently. In general, for nonlinear theories the Cauchy-Kovalevskaya theorem does not apply and difficulties may arise. In particular, for theories with nonlinear constraints a chain of such constraints could ``bifurcate'' for certain combinations of the variables (or field configurations, in field theories), and consequently, the number and/or type of constraints would change on the phase space \cite{Teite2}.

According to the Dirac's method \cite{Dirac1} we can split the constraints into two families, namely, the first-class and second-class. The first ones reflect that the theory is gauge invariant. For this type of constraints suitable additional gauge conditions can be imposed in order to remove the spurious modes or extra degrees of freedom.  For the second-class constraints we must ensure that the time evolution of the system remains on the corresponding constraint surface all the time. This can be done by means of the Dirac's method. The matrix of Poisson brackets between the second-class constraints plays a fundamental role in constructing the Dirac's brackets. These brackets can be used, together with the proper Hamiltonian, to obtain the correct time evolution that preserves the constraints. On the constraint surface the symplectic structure is encoded into the Dirac's brackets and the nondegeneracy condition falls on the matrix of Poisson brackets of the second-class constraints \cite{PreSim}. Most of the standard theories assume a constant rank for this matrix \cite{Teite}, avoiding problems of degenerate symplectic structures; however, for nonlinear theories this assumption can be broken and problems with the dynamics naturally can arise. It is worth mentioning that models with degenerate symplectic structures have been considered in the context of high energy physics, such as cosmology in brane words \cite{Z1,Z2} and gravity \cite{G2a}.    

The main goal of the present work is to analyze models that present both nonlinear second-class constraints and a spontaneous symmetry breaking. For these models we point out an apparent incompatibility between these two features that naturally leads to a degenerate symplectic structure. We explicitly show that part of the phase space where singularities arise corresponds to the minima surfaces, i.e., points where the effective Hamiltonian takes critical values and which are non trivial due to the condition of spontaneous symmetry breaking. We tackle this problem by performing the Hamiltonian analysis for theories with constraints. Following the Dirac's method, all the constraints are obtained and classified into first- and second-class. We focus on the latter, which are the ones important to construct the Dirac's brackets and to study the time evolution on the constraint surface. After that, the effective Hamiltonian together with its critical points, which include its vacua, is derived. We find that on the vacuum surfaces the Dirac's brackets are undefined due to the fact that some of the second-class constraints behave as first-class ones. This implies a modification of the number of the degrees of freedom and a potential acausal propagation near to such vacuum surfaces.  

The outline of this paper is as follows. In section \ref{Model1} we present the generic Lagrange density of the family of models we deal with. By construction, primary and secondary constraints will arise when the Hamiltonian analysis is carried out. Following the Dirac's method we identify and classify the constraints of the theory in section \ref{Hamil1}. The effective Hamiltonian as well as its critical values are derived. In section \ref{Exa1} we review some models that have been previously studied where the peculiarity we show in this work is present. We finalize in section \ref{CON1} summarizing our results and giving further concluding remarks.

\section{The model}
\label{Model1}

We focus the study on models described by the Lagrange density

\begin{equation}
\mathcal{L}=\mathcal{L}_c(q_i,\dot{q}_i,\partial_i q_j)+\rho_i A_i(q_i,\partial_i q_j)-V(q_i^2+\rho_i^2),
\label{L1a}
\end{equation}

being $q_i$, $i=1,2,..,l$, $\rho_j$, $j=1,2,...,k$, $l+k$-independent variables and $\mathcal{L}_c$, $A_i$ and $V$ functions whose arguments are properly indicated. Notice that the $\rho_i$-variables act as Lagrange multipliers and ensure the presence of primary constraints. Later we will impose as conditions a spontaneous symmetry breaking and the presence of second-class constraints and study their consequences. The absence of $\dot{\rho}_i$ is not uncommon. For theories with second-order equations of motion, criteria of covariance and a Hamiltonian with the proper sign in the kinetic term impose this characteristic, for example, in the Maxwell Lagrangian (where $\dot{A}_0=0$) and in the Einstein-Hilbert Lagrangian (see Arnowitt, Deser, and Misner (ADM) formulation) \cite{EH1,AD1,AD2}.   

 Given that the model contains constraints, due to the absence of $\dot{\rho}_i$,  the dynamics will be developed on a constraint surface. For the case of field theories such a surface can evolve in time leading to singularities similar to what occurs in the shock waves \cite{EscPot}. This was already noted in Ref. \cite{EscPot} for nonlinear electrodynamics in the Plebanski formalism, in this work we aim to generalize such studies to a broader models. In the context of bi-gravity, similar studies with second-class constraints have been performed in Refs. \cite{Iran1,Iran2}.

\section{Hamiltonian analysis}
\label{Hamil1}

In this section we delve into the Hamiltonian analysis of models described by the Lagrange density in Eq. (\ref{L1a}). As a first step we obtain the canonical Hamiltonian, $\mathcal{H}_c=\dot{q}_i\pi_{q_i}-\mathcal{L}$, given by

\begin{equation}
\mathcal{H}_c (\rho_i,q_j)=\Pi(\pi_q, \partial_j q_i)-\rho_i A_i+ V(\rho_i^2+q_i^2).
\label{Hc}
\end{equation}

The function $\Pi(\pi_q, \partial_j q_i)$, which for the moment we assume as a positive definite quantity, arises from the Legendre transformation process of $\mathcal{L}_c$ and it may include momenta related to $q_i$ and only spatial derivative terms (for construction of $\mathcal{H}_c$, this one does not include time derivatives $\partial_0$), for example

\begin{equation}
\Pi(\pi_q, \pi_X, \partial_j q_i)=\frac{1}{2}\pi_{q_i}^2+ \frac{1}{2}(\partial_j q_i)^2.
\end{equation}

The canonical algebra in the Hamiltonian approach is given by the Poisson's brackets between the coordinates (or fields) and conjugate momenta $\{q_i,\rho_i,\pi_{q_i},\pi_{\rho_i}\}$, whose non zero brackets are

\begin{equation}
\{q_i,\pi_{q}^j\}=\delta_i\,^j,\quad\quad\{\rho_i,\pi_{\rho}^j\}=\delta_i\,^j.
\end{equation}   

The definition of the conjugate canonical momenta of $\rho_j$ lead to $k$ primary constraints

\begin{equation}
\phi_j\equiv \pi_{\rho_j}\simeq0,\quad\quad j=1,2,...,k.
\end{equation} 

Here the symbol $\simeq$ means they weakly vanish, in the Dirac's terminology, i.e. they are zero on the constraint surface, but not throughout the phase space. The model can contain additional primary constraints, which we denote as

\begin{equation}
\psi_i\simeq0, \quad\quad i=1,2,...,s.
\end{equation}  

Following the Dirac's method, we write the extended Hamiltonian as follows

\begin{equation}
\mathcal{H}_{\textrm{ext}}= \mathcal{H}_c+ \lambda_i\phi_i+ \beta_j\psi_j,
\label{Hext1}
\end{equation}

where $\lambda_i$ and $\beta_j$ are Lagrange multipliers and $\mathcal{H}_c$ is defined in Eq. (\ref{Hc}). The dynamics of the $\rho_i$ variables are given by the time evolution of the quantities $\lambda_i$ since that, from the Hamilton equations, we obtain

\begin{equation}
\frac{d{\rho_i}}{dt}= \{\rho_i, \mathcal{H}_{\textrm{ext}}\}=\lambda_i.
\label{rho1234}
\end{equation}

In this way, $\dot{\rho_i}$ will be well defined as long as the quantities $\lambda_i$ have a smooth evolution, without singularities, or being fixed. We will analyze this issue later.  Continuing with the Dirac's method, in order to preserve the $s+k$ primary constraints $\{\phi_i,\psi_j\}$, we  should study their time evolution. For $\phi_i$, the following secondary constraints are produced

\begin{equation}
\Gamma_i\equiv\dot{\phi}_i=\{\phi_i,\mathcal{H}_{\textrm{ext}}\}=A_i-2\rho_i V_\xi,
\label{G12}
\end{equation}
being $V_\xi$ the derivative of $V$ respect to its argument $\xi\equiv q_i^2+\rho_i^2$. The time evolution of $\Gamma_i$ produces

\begin{equation}
\dot{\Gamma}_i=\{\Gamma_i,\mathcal{H}_{\textrm{ext}}\}= \Delta_i-2\lambda_j (\delta_{ij}  V_\xi+2V_{\xi\xi}\rho_i\rho_j),
\label{gamma1}
\end{equation}
where $ \Delta_i\equiv\{\Gamma_i, \mathcal{H}_c+\beta_j\psi_j\}$. As long as the matrix 

\begin{equation}
G_{ij}=\delta_{ij} V_\xi+2V_{\xi\xi}\rho_i\rho_j,
\label{Gij}
\end{equation}
can be inverted, Eq. (\ref{gamma1}) fixes the Lagrange multipliers $\lambda_i$ as

\begin{equation}
\lambda_i=\frac{1}{2}(G^{-1})_{ij}\Delta_j,
\end{equation}

 and no more constraints are produced from the chain of primary constraints $\phi_i$. 

The time evolution of the primary constraints $\psi_j$ may produce additional constraints, which we denote as $\Psi_j\equiv\{\psi_j,\mathcal{H}_{\textrm{ext}}\}$, with $j=1,2,...,m\leq s$ . Without loss of generality, let us assume that no more constraints appear. At this step, the Dirac's method stops and we need to classify the constrains $\{\phi_i,\Gamma_i,\psi_j, \Psi_j\}$ into first- and second-class. The $2k$ constraints $\{\phi_i,\Gamma_j\}$, or their linear combinations, do not commute between them, which means that they are second-class. Given that we did not specify completely the primary constraints $\psi_j$, it is not possible to classify the subset $\{\psi_j, \Psi_j\}$; however, we can assume that we split such constraints such that $\bar{\psi}_j$, $j=1,,2...,p$, are second-class and $\bar{\Psi}_j$, $j=1,2,...,o$ are first-class, where $o+p\leq s+m$ and the inequality holds when the set $\{\psi_i,\Psi_j\}$ is not linearly independent. The only requirement  we impose on $\bar{\psi}_j$ is that it does not contain the variables $\rho_i$. In other words, $\bar{\psi}_j$ characterizes a chain of constraints of the variables $q_i$ and its momenta, while $\{\phi_i,\Gamma_j\}$ a chain that contains both the $q_i$ and $\rho_i$ variables and momenta. The counting of degrees of freedom reveals that the model contains

\begin{eqnarray}
\# \textrm{d.o.f.}&=& 2\times\textrm{variables}- \textrm{second-class constraints}-2\times\textrm{first-class constraints} \\ \nonumber
&=&  2(l+k)-(2k+p)-2o \\ \nonumber &=& 2l-p-2o,
\end{eqnarray}  
in phase space. The first-class constraints $\bar{\Psi}_j$ generate gauge transformations and gauge fixing conditions should be imposed in order to work with the real degrees of freedom (d.o.f). Once the spurious d.o.f.'s have been removed, the dynamics of the model is performed on the constraint surface where the symplectic structure is characterized by the Dirac's brackets and the Hamiltonian subjected to the proper constraints, first-class, second-class and gauge conditions. Let us denote $\bar{\mathcal{H}}_{eff}$ as the Hamiltonian after impose all the aforementioned constraints, from Eq. (\ref{Hext1}) and (\ref{Hc}) this Hamiltonian is given by
 
 \begin{equation}
 \bar{\mathcal{H}}_{eff}=\Pi(\pi_q, \partial_j q_i)-2\rho^2 V_\xi+ V(\rho_i^2+q_i^2),
 \end{equation}
 
where the constraints in Eq. (\ref{G12}) have been strongly imposed and $\rho^2=\rho_i\rho_i$.

The relevant matrix for the Dirac's brackets arises from the Poisson brackets between the second-class constraints, and by construction, taking the order $\{\bar{\psi}_i,\phi_i,\Gamma_i\}$, has the following form

\begin{equation}
\textbf{M}=\begin{pmatrix}
\textbf{A} & \textbf{0}_a & \textbf{B}\\
\textbf{0}_b & \textbf{0}_c & \textbf{C} \\
-\textbf{B}^T & -\textbf{C}^T & \textbf{D} \\
\end{pmatrix},
\end{equation}

where the block matrices are given by

\begin{equation}
\textbf{A}=\begin{pmatrix}
 \{\bar{\psi}_i,\bar{\psi}_j\}\\
\end{pmatrix},\quad
\textbf{B}=\begin{pmatrix}
 \{\bar{\psi}_i,\Gamma_j\}\\
\end{pmatrix},\quad\quad
\textbf{C}=\begin{pmatrix}
 \{\phi_i,\Gamma_j\}\\
\end{pmatrix},\quad 
\textbf{D}=\begin{pmatrix}
\{\Gamma_i,\Gamma_j\}
\end{pmatrix}.
\end{equation}

The null matrices correspond to the Poisson brackets

\begin{equation}
\textbf{0}_a=\begin{pmatrix}
 \{\bar{\psi}_i,\phi_j\}\\
\end{pmatrix},\quad
\textbf{0}_b=\begin{pmatrix}
 \{\phi_i,\bar{\psi}_j\}\\
\end{pmatrix},\quad\quad
\textbf{0}_c=\begin{pmatrix}
 \{\phi_i,\phi_j\}\\
\end{pmatrix},
\end{equation}

The determinant of the matrix $\textbf{M}$ results

\begin{equation}
|\textrm{Det}(\textbf{M})|= \textrm{Det}(\textbf{A}) \textrm{Det}(\textbf{C})^2.
\label{Det2}
\end{equation}

The determinant of the matrix $\textbf{C}=2(\delta_{ij}V_\xi+2\rho_i\rho_jV_{\xi\xi})$ can be explicitly evaluated and is given by

\begin{equation}
|\textrm{Det}(\textbf{C})|= 2^k(V_\xi+2\rho^2V_{\xi\xi}).
\label{Det1234}
\end{equation}

 The rank of the matrix $\textbf{M}$ is not constant through all phase space. In particular, when the condition $V_\xi+2\rho^2V_{\xi\xi}=0$ holds, this matrix is not invertible. This occurs due that on the surface that defines the condition $V_\xi+2\rho^2V_{\xi\xi}=0$ some of the second-class constraints behave as first-class, i.e. they commute with all the remaining constraints.  Notice that on such a surface, the Lagrange multipliers $\lambda_i$ are not fixed by the condition in Eq. (\ref{gamma1}) since that (\textrm{Det}(\textbf{G})$\propto$\textrm{Det}(\textbf{C})). The arbitrariness of Lagrange multipliers is a characteristic of the first-class constraints. For each pair of second-class constraints that become first-class the model ``loses'' one d.o.f. on the configuration space. One might be tempted to neglect this issue by working in the regions of the phase space where the rank of the matrix $\textbf{M}$ is fixed, i.e. when Det$(\textbf{M})\neq0$; however, in the case of spontaneous symmetry breaking, such surfaces are related with the conditions of the minima and we cannot omit their existence.

 \subsection{Minima conditions}
   
The critical points of the Hamiltonian $\bar{\mathcal{H}}_{eff}$, which include the minima, are given by the following conditions
 
 \begin{eqnarray}
 \label{Cond1}
 && \frac{\partial \bar{\mathcal{H}}_{eff} }{\partial q_i}=2q_{0i} (V_\xi-2\rho^2V_{\xi\xi})=0, \\ \nonumber
&&  \frac{\partial \bar{\mathcal{H}}_{eff} }{\partial \rho_i}=-2\rho_{0i} (V_\xi+2\rho^2V_{\xi\xi})=0.
 \end{eqnarray}

The surfaces defined by the configurations $\Sigma\equiv\{q_{0i}, \rho_{0i}\}$ that satisfy (\ref{Cond1}) determine the critical points of the effective Hamiltonian. The equations (\ref{Cond1}) are trivially satisfied when $q_{0i}=0$ and $\rho_{0i}=0$. Now, if we impose a spontaneous symmetry breaking, this necessarily implies the existence of a degenerate nontrivial vacuum. It means that $q_{0i}\neq0$ and/or $\rho_{0i}\neq0$, and this leads us to the three generic cases:

\begin{itemize}
\item $\rho_{0i}=0$ and $q_{0j}\neq0$

The conditions in Eq. (\ref{Cond1}) imply that $V_\xi=0$.

\item $\rho_{0i}\neq0$ and $q_{0j}=0$

The conditions in Eq. (\ref{Cond1}) imply that $V_\xi+2\rho^2V_{\xi\xi}=0$.

\item $\rho_{0i}\neq0$ and $q_{0j}\neq0$

The conditions in Eq. (\ref{Cond1}) imply that $V_\xi=V_{\xi\xi}=0$.
\end{itemize}

For all the above cases, the determinant in Eq. (\ref{Det1234}), and therefore the determinant in Eq. (\ref{Det2}), is zero. Accordingly, for all the minima surfaces the matrix $G_{ij}$ appearing in Eq. (\ref{Gij}) is degenerate and, therefore, non invertible. 

Notice that the other way around is not true, we may stay on the degenerate surface, where the determinant in Eq. (\ref{Det2}) is zero, and out of the minima surface where the minima conditions in Eq. (\ref{Cond1}) do not hold. 

We have proved that, for these models with second-class constraints, the condition of spontaneous symmetry breaking implies the presence of vacuum surfaces where the symplectic structure, or the Dirac's brackets, becomes degenerate. This particularity may lead to singularities for some variables. In particular,  the time evolution of the $\rho_i$-variables in Eq. (\ref{rho1234}) becomes undefined (the Lagrange multipliers $\lambda_i$ cannot be properly fixed). The emergence of first-class constraints on the vacuum surface can be interpreted as an emergence of a new gauge invariance, which \textit{frozen} degrees of freedom.

\bigskip

In the former derivation we have considered a finite-dimensional system. However, the extension to infinite-dimensional field theories occurs in the same way and it does not modify the conclusion we have previously presented. We will see this in the next section, in which we review some field theories where the features and pathologies worked out in the present paper appear.

\section{Concrete examples}
\label{Exa1}

In this section we list three Lagrange densities that belong to the family of models we consider in this work. The goal will be to show that they possess the features we described in the previous section, namely, second-class constraints and a spontaneous symmetry breaking. On the minima surfaces of the effective Hamiltonian, the second-class constraints behave as first-class and the dynamics near to such surfaces suffers from a singular behaviour for some variables.    

\subsection{Bumblebee models}

The first example corresponds to a toy model to introduce spontaneous Lorentz symmetry breaking within the Standard-Model Extension framework \cite{SME1,SME2}, which is called the Bumblebee model. The corresponding Lagrange density is of the form \cite{Bumble1aa,Bumble2a2}

\begin{equation}
\mathcal{L}_B(B_\nu)= -\frac{1}{4}B_{\mu\nu}B^{\mu\nu}-\frac{\kappa}{2}V(\xi),
\label{BBN}
\end{equation}

where $B_{\mu\nu}=\partial_\mu B_\nu-\partial_\nu B_\mu$, $\xi=B^\nu B_\nu \pm b^2$, $\kappa$ is a dimensionless positive constant whereas $b^2 $ is another positive constant with dimension of $[\textrm{mass}]^2$. The potential $V$ in (\ref{BBN}) induces a nonzero vacuum expectation value defined by the relation $\langle B_0^\nu B_{0\nu}\rangle= \mp b^2$. The vector $\langle B_0^\nu\rangle$ specifies a particular direction on the spacetime and breaks the \textit{active} Lorentz transformations. It has been proved that such model mimics the standard Maxwell electrodynamics \cite{EscAlbert}, but only when special initial conditions are chosen. Such conditions force the model to remain in the vacuum where the effect of the potential disappears and the variable $B_0$, that can cause divergencies, as well as its associated conjugate momentum are zero all the time. Further studies on this model in curved and flat spacetimes can be found in Refs. \cite{bb1,bb2,bb3,bb4,bb5,Petrov1}.

The Lagrange density in Eq. (\ref{BBN}) can be cast into the form of Eq. (\ref{L1a}) identifying the variables $\rho_i$, used in the present work, with the field $B_0$ of the Bumblebee model. The canonical Hamiltonian, $\mathcal{H}_c=\dot{B}_i\Pi^i-\mathcal{L}_B$, is given by  \cite{EscAlbert}

\begin{equation}
\mathcal{H}_c=\frac{1}{2}(\Pi^j)^2+\frac{1}{4}(B_{ij})^2- B_0\partial_i \Pi^i+ \frac{\kappa}{2}V(\xi),
\end{equation}

being $\Pi^j$ the canonical momenta associated to the fields $B_j$, $j=1,2,3$. The model contains only two constraints

\begin{equation}
\phi_1=\Pi_0, \quad\quad \phi_2=\partial_j \Pi^j-\kappa B_0 V^\prime,
\end{equation}

where the prime in $V$ indicates a derivative respect to the argument $\xi$, i.e. $V_\xi=\frac{\partial V}{\partial \xi}$. The Poisson bracket between the constraints, in general, is not zero

\begin{equation}
\{\phi_1,\phi_2\}=\kappa[V'+2B_0^2V''],
\label{B1as}
\end{equation}

which tells us that they belong to the second-class type. The number of variables and constraints imply that the model contains six degrees of freedom on the phase space. The effective Hamiltonian, after setting strongly the second-class constraint $\phi_2=0$, can be rewritten as follows

\begin{equation}
\mathcal{H}_{eff}=\frac{1}{2}(\Pi^j)^2+\frac{1}{4}(B_{jk})^2-\kappa B_0^2 V'(\xi)+\frac{\kappa}{2}V(\xi).
\end{equation}

The conditions of the minima will be given by

\begin{eqnarray}
\label{minima1}
&&\frac{\partial \mathcal{H}_{eff}}{\partial B_0}=- \kappa B_0 (V'+2B_0^2V''), \\
\label{minima2}
&&\frac{\partial \mathcal{H}_{eff}}{\partial B_i}=- \kappa B^i (V'-2B_0^2V'').
\end{eqnarray}

It is straightforward to prove that any field configuration that satisfies the above conditions implies that the Poisson bracket in Eq. (\ref{B1as}) is identically zero, indicating that on such a surface the constraints $\phi_1$ and $\phi_2$ behave as first-class, modifying the number of degrees of freedom. Notice that non trivial solutions of Eqs. (\ref{minima1})-(\ref{minima2}) exist due to the spontaneous symmetry breaking, which implies a degenerate vacuum expectation value with $B_0\neq0$ and/or $B_i\neq0$. The Dirac's brackets,  which are needed to study the dynamics on the constraint surface, are not well defined on the points where the Poisson bracket in Eq. (\ref{B1as}) is zero. 

Such as we commented in the previous section, the variable $B_0$ should suffer some divergencies near to the vacuum surface. The equation of motion for $B_0$ \cite{EscAlbert} 

\begin{equation}
\dot{B}_0=\frac{1}{(V^\prime+2B_0V^{\prime\prime})}[\partial_i(B_i V^\prime)+2B_0B_iV^{\prime\prime}(\Pi^i+\partial_i B_0],
\end{equation}

reveals that such is the case in the present model. A careful analysis should be done if the dynamics or linearization, around the surface $V^\prime+2B_0V^{\prime\prime}=0$, is required. Due to the nonlinear character of the constraints some pathologies, beyond what we mentioned in the present work, can arise Ref. \cite{EscYuri}.

\subsection{Antisymmetric rank-2 tensor fields}

The generalization of the Bumblebee model was studied in Ref. \cite{Seifert1}. Here the basic variables are characterized by an antisymmetric rank-2 tensor $B_{ab}$. The Lagrange density is given by

\begin{equation}
\mathcal{L}(B_{ab})=-\frac{1}{12}F_{abc} F^{abc}- V(X),
\label{LB12}
\end{equation}

where 

\begin{equation}
F_{abc}\equiv 3\partial_{[a}B_{bc]},\quad\quad X=B_{ab} B^{ab}.
\end{equation}

Once again, upon integration by parts, the Lagrange density (\ref{LB12}) can be cast into the form of Eq. (\ref{L1a}).  With the definitions

\begin{equation}
P^i=B^{0i},\quad\quad Q^i=\frac{1}{2}\epsilon^{ijk} B^{jk},\quad\quad X=B^{ab}B_{ab}=-2\vec{P}^2+2\vec{Q}^2,
\end{equation}

and following the Dirac's method, the authors in Ref. \cite{Seifert1} derived the canonical Hamiltonian $\mathcal{H}_c=\vec{\Pi}_Q\cdot \dot{\vec{Q}}-\mathcal{L}$. It reads

\begin{equation}
\mathcal{H}_c=\frac{1}{2}\vec{\Pi}^2_Q-\vec{P}\cdot(\vec{\nabla}\times\vec{\Pi}_Q)-\frac{1}{2}(\vec{\nabla}\cdot\vec{Q})^2+V(X),
\end{equation}

and possesses six constraints

\begin{equation}
\vec{\Phi}_1=\vec{\Pi}_P,\quad\quad \vec{\Phi}_2=\vec{\nabla}\times\vec{\Pi}_Q+4V_X\vec{P}.
\end{equation}

The Poisson brackets between the constraints

\begin{equation}
\{\Phi_1^i,\Phi_2^j\}=-4(\delta^{ij}V_X-4P^i P^j V_{XX}),
\end{equation}

with determinant

\begin{equation}
\label{detC}
|\{\Phi_1^i,\Phi_2^j\}|=4^3V_X^2(V_X-4P^2 V_{XX}), \quad\quad\quad \ \,\, P^2=\vec{P}\cdot\vec{P},
\end{equation}

shows that, in general, all of them are second-class and no gauge invariance is present in this model. The counting of variables and constraints tell us that the model contains six degrees of freedom on the phase space. The effective Hamiltonian, after imposing strongly $ \vec{\Phi}_2=0$ to $\mathcal{H}_c$, takes the form

\begin{equation}
\mathcal{H}_{eff}=\frac{1}{2}\vec{\Pi}^2_Q+4P^2V_X-\frac{1}{2}(\vec{\nabla}\cdot\vec{Q})^2+V(X).
\end{equation}

The critical field configurations obey the following equations

\begin{eqnarray}
\label{minima1c}
&&\frac{\partial \mathcal{H}_{eff}}{\partial \vec{P}}=4\vec{P}(V_X-4PV_{XX}), \\
\label{minima2c}
&&\frac{\partial \mathcal{H}_{eff}}{\partial \vec{Q}}= 2\vec{Q}(V_X+4PV_{XX}).
\end{eqnarray}

Similarly to the Bumblebee model, the conditions of the minima are fulfilled solely on the surface where the determinant defined in Eq. (\ref{detC}) vanishes. It indicates that some of the second-class constraints behave as first-class. The authors in Ref. \cite{Seifert1} also concluded that the evolution between field configurations on the vacuum manifold and field configurations off the vacuum manifold is rather ill-posed, in complete agreement with our statement regarding the time evolution of the variables $\rho_i$, which in this case are identified with the variables $\vec{P}$ of the model (\ref{LB12}).

\subsection{Pleba\'nski models}

The last example corresponds to the family of the so called Pleba\'nski models. They were initially introduced to describe nonlinear electrodynamics theories \cite{Pleba1}. Recently, they have been studied in the context of black holes and gravitational waves \cite{Flores1}. The model is characterized by the Lagrange density 

\begin{equation}
\mathcal{L}(P_{\mu\nu},A_\mu)=-P^{\mu\nu}\partial_{\mu}A_\nu-V(P),
\end{equation}
where the variables are the vector potential $A_\mu$ and  the antisymmetric tensor $P_{\mu\nu}$, which are treated as independent fields. The potential $V=V(P)$ is a function of the Lorentz scalar

\begin{equation}
P=\frac{1}{4}P_{\mu\nu}P^{\mu\nu}.
\end{equation}

Thus, this model preserves gauge invariance from the beginning. In Refs. \cite{EscPot,EscPot2} the authors studied the Pleba\'nski model and showed that a degenerate behavior can arise for special non constant field configurations, in agreement with the main point of the present consideration. 

From the Hamiltonian analysis one gets the canonical Hamiltonian \cite{EscPot2} 

\begin{equation}
\mathcal{H}_c= (\partial_i A_j) P^{ij} + V(P),
\label{H-E}
\end{equation}

and all the constraints. Particularly, the second-class constraints turn out to be 

\begin{align}
\label{Theta33}
\Theta^3_i &=  \Pi_i^A \approx 0\>,\\ 
\Theta^4_i &=  \pi_i-A_i \approx 0\>,\\
\Theta^5_{ij} &= \pi_{ij} \approx 0\>,\\
\Theta^6_{ij} &= (\partial_i A_j-\partial_j A_i)+ V_P P_{ij}\approx 0\>.
\label{Theta-6}
\end{align} 

being $\Pi_i^A,\, \pi_i$ and $\pi_{ij}$ the canonical momenta of $A_i,\, P_{0i}$ and $P_{ij}$, respectively. With the definitions

\begin{eqnarray}
D_i &&\equiv P_{0i} = (D_x, D_y, D_z)\>, \\
H_i &&\equiv \tilde P_{0i} = \frac{1}{2}\epsilon_{0ijk}P^{jk}= (H_x, H_y, H_z)\>, \\
\label{field-defs}
\end{eqnarray}

the determinant of the Poisson brackets between the constraints (\ref{Theta33})-(\ref{Theta-6}) is

\begin{align}
\left|\left\{\Theta^i(x),\Theta^j(y)\right\}_{x^0=y^0}\right| &=
V_P^3 (V_P+H^2V_{PP})\delta^3(\vec{x}-\vec{y})\>, \quad\quad (H^2=\vec{H}\cdot\vec{H}),
\label{detB}
\end{align}  

showing that they are effectively second-class. After substituting the second-class constraint, appearing in Eq. (\ref{Theta-6}), into the canonical Hamiltonian $\mathcal{H}_c$ the effective Hamiltonian results in

\begin{equation}
\mathcal{H}_{eff} = -H^2 V_P  + V(P)\>.
\label{effective-Hamiltonian}
\end{equation}

The conditions for the minima are given by

\begin{eqnarray}
\label{mini1}
&&\frac{\partial \mathcal{H}_{eff}}{\partial H_i} = -H_i(V_P + H^2 V_{PP}), \\ 
\label{minima2ba}
&&\frac{\partial\mathcal{H}_{eff}}{\partial D_i} = -D_i(V_P - H^2 V_{PP}).
\end{eqnarray}

Once again, the above conditions are satisfied on a surface where the determinant  (\ref{detB}) is zero. Here the variables $\rho_i$ in the present work, are identified with the variables $\vec{H}$ of the Plebanski model. For the time evolution we arrive at the equation \cite{EscPot2}

\begin{equation}
\dot{\vec{H}} = \dfrac{1}{V_P}\left(-\vec{\nabla}\times(V_P\vec{D})
+ \dfrac{N\,V_{PP}\,\vec{H}}{V_P + H^2V_{PP}}\right)\>,
\label{dot-H}
\end{equation}
where
\begin{equation}
N \equiv \vec{H}\cdot(\vec{\nabla}\times V_P\vec{D}) + V_P\vec{D}\cdot(\vec{\nabla}\times\vec{H})\>.
\label{N}
\end{equation}

The Eq. (\ref{dot-H}) shows that the model can present an acausal propagation near to the vacuum surface, where the denominator in the second term blows up.

Therefore, for the previous three models we find the same characteristics and features. The symplectic structure, or Dirac's brackets, becomes degenerate on the surfaces where the effective Hamiltonian takes critical values (including its minima). This is because second-class constraints behave as first-class on this surface. As a physical consequence, there is a change in the number of degrees of freedom and an acausal propagation appears.   

\section{Conclusions}
\label{CON1}

In the present work we have studied a family of models that present both nonlinear second-class constraints and spontaneous symmetry breaking (SSB). We have explicitly shown that the aforementioned properties imply that the symplectic structure becomes degenerate on the vacuum surfaces. We deal with the problem by performing a Hamiltonian analysis for theories with constraints. Following the Dirac's method, we found and classified all the constraints of the model. After imposing strongly the second-class constraints we derived the effective Hamiltonian and obtained its critical points. In these nonlinear theories, or models with second-class constraints, the relevant quantity to determine the critical points is the effective Hamiltonian and not only the potential. We explicitly showed that, assuming a SSB, some of the second-class constraints behave as first-class on the vacuum surfaces. 

The symplectic structure on the constrained surface and the time evolution of the system are codified by the Dirac's brackets. The latter are not well defined on the vacuum surface since that they are constructed from the matrix of the Poisson brackets between the second-class constraints. This can lead to divergencies in the time evolution of some variables or fields. Reviewing three particular models, we explicitly emphasize all these features and discuss the potential inconsistencies that may appear.

%When the mechanism of SSB is triggered by a potential, this should be nonlinear, in order to produce non trivial vacua. If the theory contains second-class constraints this can lead us to difficulties to describe the 

One of the main problems in these models is the dynamical evolution near the vacua. Generally, one is interested in studying small perturbations around the vacuum state, employing a linearization method. However, for the models we consider it is not clear that such a process can be done in a consistent manner, namely, avoiding problems of causality, such as it is shown in Ref. \cite{EscPot}. The dynamics and/or constraints of the models do not prevent that an initial configuration out of the vacuum state evolves to it, and once this happens it is not clear how it would evolve afterwards. 

If  the full theory is nonlinear, it is likely to be qualitatively different from the linearized  theory in some aspects, for example in the number and type of constraints. As a consequence, the number of the degrees of freedom can change through the phase space. The particular importance is that the surface where we find this phenomenon is precisely the vacuum surfaces, and therefore we cannot just ignore it and work in a different part of the phase space. 

In view of the difficulties we have noted, some alternatives can be considered. In order to avoid part of the pathologies, one could simply constrain the model on the vacuum surface via a Lagrange multiplier with a term of the form $\lambda \,S$, where $S$ is a quantity that defines the vacuum surface and $\lambda$ the multiplier. The equation of motion for $\lambda$ leads us to the condition $S=0$ all the time, and the problems with the dynamical evolution near such a surface would be avoided. This alternative should be further investigated in detail given that the introduction of the Lagrange multiplier modifies the phase space as well as the algebra of the primary constraints in significant ways \cite{S23}. A different alternative would be to determine if there exist special initial conditions that guarantee that the system does not evolve to the problematic surfaces. This would be a tough task since we cannot use the standard mathematical tools, such as the Cauchy-Kovalevskaya theorem given that its hypotheses do not hold in the present nonlinear models.

It would be interesting to extend the present study to more general models. Specifically, to investigate if second-class constraints and spontaneous symmetry breaking always imply a degenerate symplectic structure (on the vacuum surface) and if an acausal propagation for some modes is present.

%It is worth to mention that 

\acknowledgements

C. A. E. is partially supported by the Project PAPIIT No. IN109321.

\appendix

\end{document}